# Shell DFT-1/2 method towards engineering accuracy for semiconductors: GGA versus LDA

Hanli Cui,[1] Shengxin Yang,[1] Jun-Hui Yuan,[1,*] Li-Heng Li,[1] Fan Ye,[1] Jinhai Huang,[1] Kan-Hao Xue,[1,2,*] and Xiangshui Miao[1,2]

[1] School of Integrated Circuits, School of Optical and Electronic Information, Huazhong University of Science and Technology, Wuhan, 430074, China

[2] Hubei Yangtze Memory Laboratories, Wuhan 430205, China

*Corresponding Authors, Email: xkh@hust.edu.cn (K.-H. Xue); yuanjh90@hust.edu.cn (J.-H. Yuan)

## Abstract

The Kohn-Sham gaps of density functional theory (DFT) obtained in terms of local density approximation (LDA) or generalized gradient approximation (GGA) cannot be directly linked to the fundamental gaps of semiconductors, but in engineering there is a strong demand to match them through certain rectification methods. Shell DFT-1/2 (shDFT-1/2), as a variant of DFT-1/2, is a potential candidate to yield much improved band gaps for covalent semiconductors, but its accuracy depends on the LDA/GGA ground state, including optimized lattice parameters, basic Kohn-Sham gap before self-energy correction and the amount of self-energy correction that is specific to the exchange-correlation (XC) functional. In this work, we test the LDA/GGA as well as shDFT-1/2 results of six technically important covalent semiconductors Si, Ge, GaN, GaP, GaAs and GaSb, with an additional ionic insulator LiF for comparison. The impact of XC flavor (LDA, PBEsol, PBE and RPBE), either directly on the gap value, or indirectly through the optimized lattice constant, is examined comprehensively. Moreover, we test the impact of XC flavor on LDA/GGA and shDFT-1/2 gaps under the condition of fixed experimental lattice constants. In-depth analysis reveals the rule of reaching the best accuracy in calculating the electronic band structures of typical covalent semiconductors. Relevant parameters like lattice constant, self-consistency in shDFT-1/2 runs, as well as the exchange enhancement factor of GGA, are discussed in details.

# 1. Introduction

Compared with metal and alloy research where density functional theory (DFT) has played a vital role, in semiconductor-related calculations and simulations, non-first-principles methods such as tight binding [1] and $k \cdot p$ perturbation [2] have in general dominated the literature. The reasons may be two folds. On the one hand, simulation at the semiconductor device level usually requires ultra-large supercells, and the computational load of DFT even at its local density approximation (LDA) level is too high. Nevertheless, as computers are unceasingly getting stronger, and the feature dimension of microelectronic devices is scaling-down, the computational complexity problem of first-principles methods is greatly alleviated. On the other hand, LDA fails to recover the correct band gap value, which is however the most important parameter for semiconductors. Typical LDA band gaps are ~40% smaller than experimental [3], and the slightly more advanced generalized gradient approximation (GGA) technique does not improve upon LDA. More computationally demanding methods like hybrid functionals [4,5] as well as the quasi-particle approach within the *GW* approximation [6,7] are substantially better, but their computational loads are 2-3 orders of magnitude higher than LDA. Hence, to recover satisfactory electronic band structures for semiconductors at the LDA complexity, especially for the band gaps, is a key task for DFT to enter semiconductor research and semiconductor industry.

Some state-of-the-art meta-GGA functionals, such as SCAN [8] and TB09 [9], already perform much better than LDA/GGA, with computational complexity lying in between GGA and hybrid functionals, and much less demanding than hybrid functionals. However, their comprehensive performances in terms of speed, convergence, and more importantly the band gap accuracy for semiconductors, still do not fit the stringent engineering accuracy. Another approach is to use some post-correction methods, which may only rectify the electronic structure, rather than the total energy. This relaxes the constraints on the rectification method, and one may obtain accurate and efficient methods for band gap calculation. The DFT-1/2 method as proposed by Ferreira *et al.* [10], as well as its extension shell DFT-1/2 (shDFT-1/2 for short) and shell DFT-*x*-*y* (shDFT-*x*-*y* for short) for covalent semiconductors [11], are such techniques. Among them, shDFT-*x*-*y* allows for the fitting of band gap to experimental value through an adjustable parameter $x$ (or $y$, since $x + y = 1/2$).

However, from the *ab initio* sense, empirical-parameter-free shDFT-1/2 enables first-principles prediction of band gaps for semiconductors and insulators, with band gap mismatch typically limited to 0.5 eV from experimental [11]. The computational complexity of DFT-1/2 and shDFT-1/2 is comparable to LDA/GGA when applied to supercells, with <15% more computer time [11]. Nevertheless, such accuracy still does not fit the engineering need in microelectronics, and LDA usually underestimates the lattice constants for solids. The lattice constant mismatch inevitably causes certain band gap inaccuracy. One could naturally turn to shGGA-1/2, as several GGA functionals such as AM05 [12], Wu-Cohen [13] and PBEsol [14] are well-known to predict lattice constants very similar to experimental. A previous research reveals that PBEsol performs best in shGGA-1/2 for wide-gap oxides $Al_2O_3$, $TiO_2$, $ZrO_2$ and $HfO_2$ [15]. However, that work did not cover the covalent semiconductors, and for small-gap semiconductors the engineering requirement for band gap accuracy is even more stringent. In this work we explore, both from theoretical and computational, the influence of GGA exchange-correlation (XC) on the shDFT-1/2 results compared with LDA XC, for several technically important semiconductors.

## 2. Theoretical and computational

### 2.1 DFT-1/2 and shDFT-1/2

The DFT-1/2 method is a modern DFT version of Slater's half-occupation technique [16], which had been used with the X$\alpha$ method for ionization energy calculations in atoms and molecules [17]. In extending it from molecules/clusters to extended systems, Ferreira *et al.* [10] started from the Slater-Janak theorem [18], with some further assumptions such as the linear variation of Kohn-Sham energy eigenvalue with respect to the occupation of that orbital. Hence, -1/2 means a half way between the initial ground state and the ionized state, *i.e.*, the semiconductor with one hole. This makes sense because the Kohn-Sham eigenvalues only characterize the energies for fixed-electron systems, namely those with the same electron number ($N$) as the neutral ground state. Nevertheless, the true physical ionization process involves both $N$-electron and ($N$-1)-electron systems. It is the -1/2 correction that could correctly account for the discrepancy in electron numbers. Regarding the electron affinity, the same argument holds, but usually the conduction band (CB) electron occupies a Bloch-like state with nearly-null self-energy, thus the +1/2 CB correction [19] can usually be neglected in most cases.

The remarkable efficiency of DFT-1/2 stems from its extremely simple way of self-energy correction, by transferring a "self-energy potential" from atomic calculations to the solid. Not only does one avoid calculating the one-particle Green's function, but the atomic pseudopotential plus the self-energy potential could be treated as an ordinary pseudopotential ready for electronic structure calculations in any specific code. Nevertheless, the long-range self-energy potential must be properly trimmed before introducing to solids, thus a crucial part of DFT-1/2 method is the cutoff function, which restricts the coverage of atomic self-energy potential through a cutoff radius $r_{cut}$ [10]. It is not an empirical parameter, but ought to be obtained variationally to render the band gap at extreme. In covalent semiconductors, it was found that a single cutoff radius is usually insufficient, and two cutoff radii ($r_{in}$ and $r_{out}$) have been introduced in shDFT-1/2. Their proper combination should yield the maximum band gap. A covalent semiconductor is usually signified by a non-zero inner cutoff ($r_{in} > 0$), and extensive tests have shown the better performance of shDFT-1/2 in IV, III-V and II-VI semiconductors [11].

In the derivation of DFT-1/2 (or similarly, shDFT-1/2), one has to assume a local or semi-local form for the XC energy. Even though the original formalism was obtained using LDA, the DFT-1/2 technique also holds for GGA. In another work by Ferreira *et al.* [20], it was claimed that the Slater-Janak theorem "is likely to be also valid for the GGA, though GGA may come in very different formulations". Indeed, GGA has intricate differences compared with LDA. The XC functional of LDA (actually its correlation part) almost exclusively comes from the quantum Monte Carlo simulation results on uniform electron gases [21]. This prototype does not possess any electron density gradient. For molecular systems, it is well-known that an exchange enhancement must be considered for finite density gradients [22]. And in solids, LDA suffers from the over-binding problem and usually predicts too small lattice constants. Through an exchange enhancement factor for finite density gradients, GGA enlarges the lattice constants. However, the Perdew–Burke–Ernzerhof (PBE) functional typically overcorrects and yields too large lattice constants, as it was mainly designed for molecular calculations. As claimed by Perdew *et al.* [14], no GGA can do both as to yield accurate total energy as well as good lattice constants for solids. So far as semiconductors are concerned, the electronic band structure is more important than the total energy, thus several GGA flavors aiming at solids can be adopted, such as AM05 [12], Wu-Cohen [13], PBEsol [14] as well as SOGGA [23]. Their common feature lies in that the exchange enhancement factor $F_X$ is less than that of PBE, which avoids predicting too large primitive cells.

Hence, the exchange enhancement function $F_X$ is a core characteristic of a particular GGA functional. In this work, we focus on its influence on the GGA as well as shGGA-1/2 band gaps for semiconductors. Hence, we select three GGA flavors to be compared with LDA, including PBEsol [14], PBE [24], and RPBE . Their $F_X$'s are getting stronger following this sequence, as illustrated in **Figure 1**. For the materials testing set, technically important covalent semiconductors in diamond or zinc blende structure—Si, Ge, GaN, GaP, GaAs and GaSb are included, with an additional ionic insulator for comparison, *i.e.*, LiF in the rock salt structure.

## 2.2 Computational settings

DFT calculations were carried out using the plane-wave-based Vienna *Ab initio* Simulation Package (VASP) [25,26], with a fixed 600 eV plane-wave kinetic energy cutoff. The LDA functional was in the form of Ceperley-Alder (CA) and parameterized by Perdew and Zunger [27]. Three distinct flavors for the GGA XC functional were considered, including PBE [42], PBE developed for solids (PBEsol) [34], and the RPBE functional developed by Hammer, Hansen and Nørskov [28]. The Monkhorst-Pack *k*-point grid [29] for Brillouin zone sampling during geometry optimization and ground state charge calculations was $15 \times 15 \times 15$, centered at the $\Gamma$ point. The projector augmented-wave method was adopted [30,31], and valence electrons were: 1s and 2s for Li; 2s and 2p for C; 2s and 2p for N; 2s and 2p for F; 3s and 3p for Si; 3s and 3p for P; 4s and 4p for Ga; 4s and 4p for Ge; 4s and 4p for As; 5s and 5p for Sb. Spin-orbit coupling was turned on during electronic structure calculations, except for LiF and GaN. Shell DFT-1/2 self-energy correction was carried out in the following manner: -1/4 e correction (or simply -1/4) for Si and Ge; -1/2 for N in GaN; -1/2 for P in GaP; -1/4 for Ga and -1/4 for As in GaAs; -1/4 for Ga and -1/4 for Sb in GaSb; -1/2 for F in LiF. The self-energy potentials were derived using a modified ATOM code [15,32].

# 3. Results and Discussion

## 3.1 Band gaps based on LDA and GGA

The delicate difference between Kohn-Sham gaps under LDA and under GGA has been paid less attention, as both severely deviate from the experimental value. It does not make sense striving to improve the gap mismatch from, say, 48% to 42%, but in case a proper rectification method exists, to improve the mismatch from 10% to 4% will be a great benefit for engineering. As shDFT-1/2 is based on the LDA/GGA ground state, it becomes worthwhile to understand the influence of XC flavor, and other factors such as lattice constant mismatch, on the Kohn-Sham gap values.

The optimized lattice constants ($a$) for the seven materials show a consistent trend, with $a_{LDA} < a_{PBEsol} < a_{PBE} < a_{RPBE}$, as seen from **Table I**. This fully agrees with the $F_X$ function for each XC functional (see **Figure 1**), where RPBE and PBEsol are GGA XCs with stronger and weaker exchange enhancement effects than PBE. The strength of exchange is therefore LDA<PBEsol<PBE<RPBE, consistent with the trend of lattice constant. The increased exchange effect tends to pull atoms/ions farther from each other, and it ought to play a role in the band gap results as well.

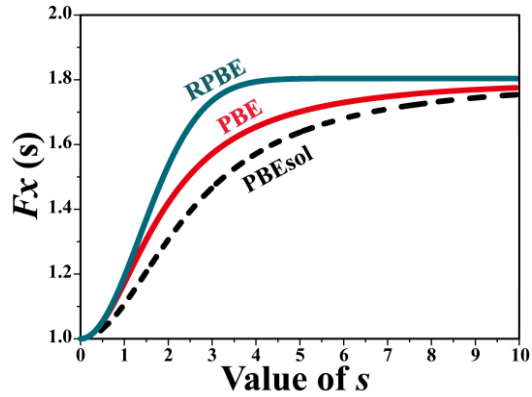

**Figure 1**. $F_X(s)$ as a function of reduced density gradient ($s$) for PBEsol, PBE and RPBE functionals.

**Table I**. LDA/GGA and shDFT-1/2 calculation results for six semiconductors (Si, Ge, GaN, GaP, GaAs, GaSb) and an ionic insulator (LiF). Units are: eV for $E_g$ and $\Delta E_g$, Å for $a$, bohr for $r_{in}$ and $r_{out}$.

|  |  |  | LDA/GGA | | | | ShDFT-1/2 | | | | |
| --- | --- | --- | --- | --- | --- | --- | --- | --- | --- | --- | --- |
|  |  |  | CA | PBEsol | PBE | RPBE |  | CA | PBEsol | PBE | RPBE |
| **Si** $E_g$ = 1.17 | Opt. | $E_g$ | 0.429 | 0.447 | 0.596 | 0.689 | $E_g$ | 1.164 | 1.213 | 1.378 | 1.500 |
|  |  | $\Delta E_g$ | **0.735** | **0.766** | **0.782** | **0.811** | $r_{in}$ | 0.7 | 0.7 | 0.7 | 0.7 |
|  |  | $a$ | 5.403 | 5.408 | 5.470 | 5.501 | $r_{out}$ | 3.1 | 3.1 | 3.2 | 3.2 |
|  | Exp. | $E_g$ | 0.455 | 0.442 | 0.557 | 0.616 | $E_g$ | 1.189 | 1.208 | 1.341 | 1.432 |
|  |  | $\Delta E_g$ | **0.734** | **0.766** | **0.784** | **0.816** | Cutoff | $r_{in}$ = 0.7, $r_{out}$ = 3.1 | | | |
|  |  | $a$ | 5.431 | | | | | | | | |
| **Ge** $E_g$ = 0.74 | Opt. | $E_g$ | 0 | 0 | 0 | 0 | $E_g$ | 0.801 | 0.757 | 0.450 | 0.293 |
|  |  | $\Delta E_g$ | **0.801** | **0.757** | **0.450** | **0.293** | $r_{in}$ | 1.5 | 1.6 | 1.8 | 1.8 |
|  |  | $a$ | 5.644 | 5.701 | 5.782 | 5.836 | $r_{out}$ | 3.3 | 3.5 | 3.5 | 3.5 |
|  | Exp. | $E_g$ | 0 | 0 | 0 | 0 | $E_g$ | 0.776 | 0.847 | 0.911 | 0.997 |
|  |  | $\Delta E_g$ | **0.776** | **0.847** | **0.911** | **0.997** | $r_{in}$ | 1.5 | 1.5 | 1.5 | 1.5 |
|  |  | $a$ | 5.664 | | | | $r_{out}$ | 3.4 | 3.3 | 3.3 | 3.3 |
| **GaN** $E_g$ = 3.51 | Opt. | $E_g$ | 1.903 | 1.629 | 1.448 | 1.381 | $E_g$ | 3.764 | 3.550 | 3.450 | 3.385 |
|  |  | $\Delta E_g$ | **1.861** | **1.921** | **2.002** | **2.004** | Cutoff | N: $r_{in}$ = 1.0, $r_{out}$ = 2.5 | | | |
|  |  | $a$ | 4.500 | 4.545 | 4.588 | 4.620 |  | | | | |
|  | Exp. | $E_g$ | 1.768 | 1.691 | 1.737 | 1.765 | $E_g$ | 3.622 | 3.616 | 3.716 | 3.799 |
|  |  | $\Delta E_g$ | **1.854** | **1.925** | **1.979** | **2.034`** | Cutoff | N: $r_{in}$ = 1.0, $r_{out}$ = 2.5 | | | |
|  |  | $a$ | 4.531 | | | | | | | | |
| **GaP** $E_g$ = 2.34 | Opt. | $E_g$ | 1.410 | 1.474 | 1.505 | 1.415 | $E_g$ | 2.655 | 2.763 | 2.775 | 2.812 |
|  |  | $\Delta E_g$ | **1.245** | **1.289** | **1.270** | **1.397** | $r_{in}$ | 0.6 | 0.6 | 0.9 | 0.9 |
|  |  | $a$ | 5.422 | 5.470 | 5.531 | 5.578 | $r_{out}$ | 2.8 | 2.9 | 3.0 | 3.0 |
|  | Exp. | $E_g$ | 1.437 | 1.456 | 1.587 | 1.681 | $E_g$ | 2.671 | 2.754 | 2.883 | 2.995 |
|  |  | $\Delta E_g$ | **1.234** | **1.298** | **1.296** | **1.314** | $r_{in}$ | 0.6 | 0.6 | 0.6 | 0.6 |
|  |  | $a$ | 5.451 | | | | $r_{out}$ | 2.9 | 2.9 | 2.9 | 2.8 |
| **GaAs** $E_g$ = 1.52 | Opt. | $E_g$ | 0.413 | 0.306 | 0.054 | 0 | $E_g$ | 1.541 | 1.820 | 1.204 | 2.287 |
|  |  | $\Delta E_g$ | **1.128** | **1.514** | **1.150** | **2.287** | $r_{in}$ | Ga: 2.1 | Ga: 2.1 | Ga: 2.2 | Ga: 2.2 |
|  |  |  |  |  |  |  |  | As: 1.4 | As: 1.4 | As: 1.4 | As: 1.5 |
|  |  | $a$ | 5.621 | 5.678 | 5.759 | 5.815 | $r_{out}$ | Ga: 3.9 | Ga: 3.9 | Ga: 4.0 | Ga: 4.1 |
|  |  |  |  |  |  |  |  | As: 3.2 | As: 3.3 | As: 3.3 | As: 3.3 |
|  | Exp. | $E_g$ | 0.288 | 0.408 | 0.471 | 0.583 | $E_g$ | 1.413 | 1.557 | 1.633 | 1.760 |
|  |  | $\Delta E_g$ | **1.125** | **1.149** | **1.162** | **1.177** | $r_{in}$ | Ga: 2.1 | | | |

| | | | | | | | | | | |
|---|---|---|---|---|---|---|---|---|---|---|
| | | | | | | | | As: 1.4 | | |
| | | $a$ | 5.653 | | | | $r_{out}$ | Ga: 3.9 | | |
| | | | | | | | | As: 3.2 | | |
| **GaSb** $E_g =$ 0.81 | Opt. | $E_g$ | 0 | 0 | 0 | 0 | $E_g$ | 0.942 | 0.871 | 0.583 | 0.442 |
| | | $\Delta E_g$ | **0.942** | **0.871** | **0.583** | **0.442** | $r_{in}$ | Ga: 2.1 | Ga: 2.1 | Ga: 2.2 | Ga: 2.2 |
| | | | | | | | | Sb: 1.6 | Sb: 1.8 | Sb: 1.8 | Sb: 1.8 |
| | | $a$ | 6.058 | 6.121 | 6.218 | 6.286 | $r_{out}$ | Ga: 3.8 | Ga: 3.9 | Ga: 4.0 | Ga: 4.1 |
| | | | | | | | | Sb: 3.6 | Sb: 3.6 | Sb: 3.6 | Sb: 3.7 |
| | Exp. | $E_g$ | 0 | 0 | 0 | 0 | $E_g$ | 0.830 | 0.965 | 1.020 | 1.102 |
| | | $\Delta E_g$ | **0.830** | **0.965** | **1.020** | **1..102** | $r_{in}$ | Ga: 2.1 | | | |
| | | | | | | | | Sb: 1.8 | Sb: 1.7 | Sb: 1.8 | Sb: 1.8 |
| | | $a$ | 6.095 | | | | $r_{out}$ | Ga: 3.9 | Ga: 4.0 | Ga: 3.9 | Ga: 4.0 |
| | | | | | | | | Sb: 3.6 | | | |
| **LiF** $E_g =$ 14.2 | Opt. | $E_g$ | 9.680 | 9.034 | 8.853 | 8.612 | $E_g$ | 13.031 | 12.630 | 12.365 | 12.244 |
| | | $\Delta E_g$ | **3.351** | **3.596** | **3.512** | **3.632** | Cutoff | F: $r_{in} = 0.4$, $r_{out} = 2.0$ | | | |
| | | $a$ | 3.907 | 4.004 | 4.060 | 4.138 | | | | | |
| | Exp. | $E_g$ | 9.240 | 9.302 | 9.477 | 9.721 | $E_g$ | 12.595 | 12.874 | 13.019 | 13.395 |
| | | $\Delta E_g$ | **3.355** | **3.572** | **3.542** | **3.674** | Cutoff | F: $r_{in} = 0.4$, $r_{out} = 2.0$ | | | |
| | | $a$ | 3.964 | | | | | | | | |

Before discussing the direct impact of $F_X$ on shDFT-1/2 calculation results, it is worthwhile to first examine its indirect impact on band gaps, through the lattice constant distinction. It is at this juncture uncertain whether the direct impact or the indirect impact should play a more important role. Hence, we carried out systematic investigation on the lattice constant-dependent band gaps as predicted by LDA or GGA, with emphasis on the LDA results because of its relatively fixed XC functional form.

Take LiF and Si as typical examples for ionic and covalent semiconductors, respectively. For a reasonable range of lattice constants, the LDA band gap for LiF decreases monotonously upon enlarging the lattice, as shown in **Figure 2(a)**. For convenience, such trend will be called negative $E_g$—$a$ correlation. LDA calculations at fixed experimental lattice constant yield smaller band gap than using LDA-optimized lattice constant, because the latter is less than the former due to the over-binding tendency of LDA. Nevertheless, in the covalent semiconductor Si, the LDA band gap exhibits a maximum at some particular lattice constant (**Figure 2(b)**). Moreover, near the experimental or LDA-optimized lattice constants, the band gap actually increases upon enlarging the lattice, demonstrating a positive $E_g$—$a$ correlation. This implies that if the lattice constant is fixed to experimental value, one can obtain a larger band gap than using the equilibrium LDA lattice constant for Si.

The trend observed in Si is not universal among covalent semiconductors. Simply put, even if a maximum could exist, the experimental lattice constant may reside on either the up-hill side or the down-hill side. We find that most covalent semiconductors follow the negative $E_g$—$a$ correlation (*i.e.*, downhill) like LiF. In diamond carbon, for instance, the band gap keeps on decreasing when the lattice constant is setting to larger values, without observing a maximum point within a sufficient range, as illustrated in **Figure 2(c)**. For GaAs, result in **Figure 2(d)** even shows a nearly linear decay of LDA band gap when the lattice is enlarged. The general trend of band gap versus lattice constant has been attributed to the effect of deformation potential [33].

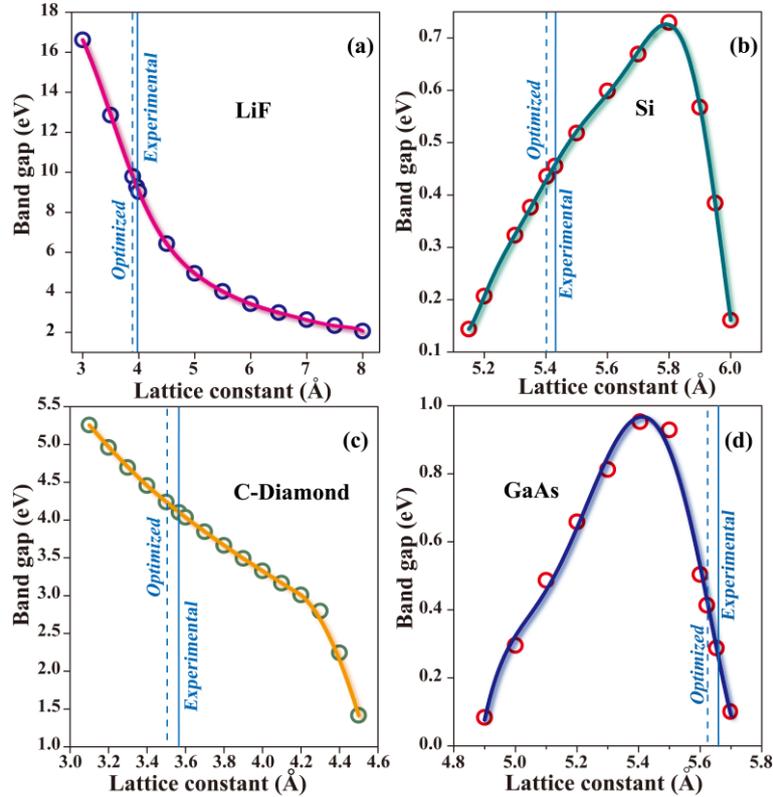

**Figure 2**. Relation of LDA band gap with respect to lattice constant for (a) LiF; (b) Si; (c) Diamond carbon; (d) GaAs.

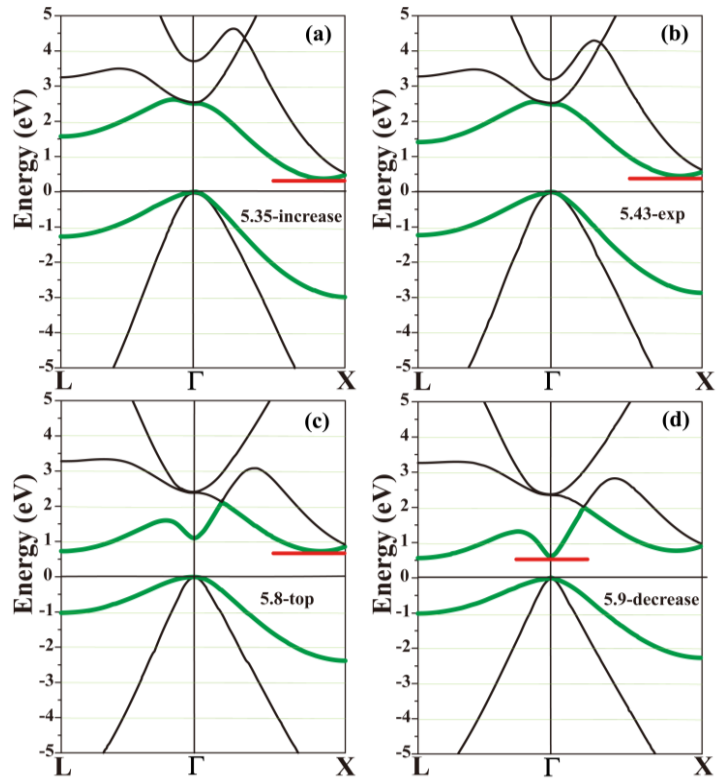

**Figure 3**. Impact of lattice constant on the LDA band structures of Si. (a) $a = 5.35$ Å, less than experimental value; (b) $a = 5.43$ Å, at experimental value; (c) $a = 5.80$ Å, roughly corresponding to the maximum band gap; (d) $a = 5.90$ Å, residing at the down-hill side.

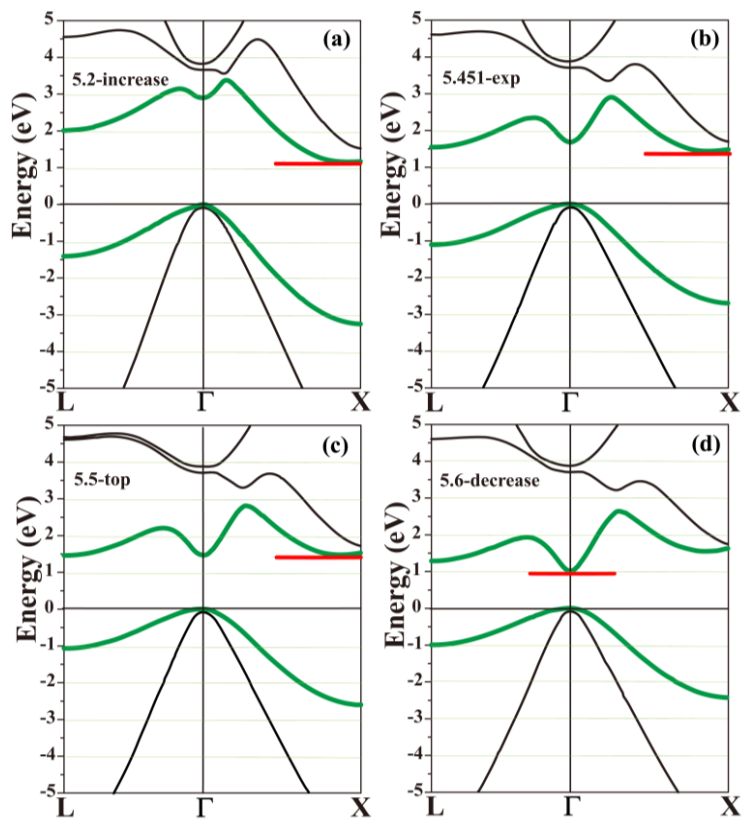

**Figure 4**. Impact of lattice constant on the LDA band structures of GaP. (a) $a = 5.20$ Å, less than experimental value; (b) *a* = 5.451 Å, at experimental value; (c) *a* = 5.50 Å, roughly corresponding to the maximum band gap; (d) *a* = 5.60 Å, residing at the down-hill side.

**Figure 3** demonstrates the energy band diagrams of Si calculated at several selected lattice constants. The direct gap feature is found to be the origin of the gap maximum, as the Γ-Γ direct gap shows a normal reduction when the primitive cell is enlarged. However, the indirect gap varies in the opposite direction, and at *a* = 5.50 Å the two kinds of gap value almost coincide, corresponding to the maximum band gap. At *a* = 5.60 Å, Si already becomes a direct gap semiconductor, though of course this is only of theoretical value as such violent stretching is rather difficult to be realized in experiments. Another indirect-gap semiconductor GaP demonstrates a similar trend as Si, and its energy band diagrams at several lattice constants are given in **Figure 4**. Although the conduction band energy at Γ is not far from that of L and near X (**Figure 4(b)**), it becomes abnormally high and the lowest edge of the band, as the lattice constant shrinks (**Figure 4(a)**) or increases (**Figure 4(d)**).

On the other hand, GGA usually predicts larger lattice constants, or similar to experimental. To explore whether the GGA band gap is in general smaller or larger than LDA, one faces a dilemma whether the comparison should be carried out using the same lattice constant, or individually-optimized lattice constant for each. In the former case, the stress levels are distinct for LDA and GGA, but for the latter case the band gaps ought to be different regardless of the XC used. However, provided that $E_g$ is highly sensitive to *a*, then the second approach will make little sense. Therefore, it is best to first fix the lattice constant to the experimental value in order to examine the impact of XC flavor on band gap values. Except for GaN, all GGA band gap values at fixed experimental lattice constants are slightly larger than that of LDA in our test set. However, at optimized lattice constants, GGA band gaps are constantly lower than that of LDA, except for Si (obvious) and GaP (to a less extent). This is consistent with the observation that Si possesses an abnormal trend of positive $E_g$—*a* correlation. In other words, these comparisons indicate that the impact of lattice constant is more significant than the choice of XC flavor.

It is worthwhile to digress into why GGA band gaps are usually larger than that of LDA when both calculated using the experimental lattice constant. A key feature of GGA against LDA lies in that a finite density gradient renders an exchange enhancement (factor $F_X$), which tends to pull down energy bands. Since the top of VB usually corresponds to higher electron density ($n$) regions, the same exchange enhancement would move the VB more than the CB, because the exchange itself is monotonous with respect to $n$ [34]. This explains why fundamentally, GGA band gaps tend to be larger than that of LDA, if sticking to experimental lattice. Yet, the enhancement factor $F_X$ does not directly depend on $\vec{\nabla} n$, but on a reduced gradient [34]

$$s = s_1 = \frac{|\vec{\nabla} n|}{2 \left(\frac{2\pi}{3}\right)^{\frac{1}{3}} n^{\frac{4}{3}}}$$

Therefore, we analyzed the distribution of $s$ in these semiconductors and insulators. For the ionic insulator LiF, we discover a clear trend of $E_g$(LDA) < $E_g$(PBEsol) < $E_g$(PBE) < $E_g$(RPBE) when fixed to experimental lattice constant. The hole distributes around the F anions, demonstrating a strong overlap with part of the high-$s$ regions (comparing **Figures 5(a)** and **5(b)**). On the other hand, in **Figure 5(c)** the shell regions near Li consists of an inner yellow (positive) region and an outer cyan (negative) region, representing a shift of electron location towards the Li core. Hence, the CB electron is most probably found within the low-$s$ regions far from the ionic cores. GGA tends to pull down the valence band (VB), but not so much for the CB, thus enlarging the band gap of LiF. Nevertheless, the high-$s$ region is quite different in covalent semiconductors, as it usually points to the gap space far from the atoms (see **Figures 5(d)** and **5(g)**), not forming shell regions surrounding the cores. On the other hand, in III-V compounds the VB hole is localized around the V-elements. However, the CB electron distribution shows delicate differences among the materials. For instance, in GaAs the CB electron is most probably found at the near-core regions for both Ga and As (**Figures 5(f)**), thus the $F_X$ function in GGA does not impact the CB of GaAs much. In GaN, nevertheless, the CB electron extends to Ga-Ga bond regions (**Figures 5(i)**). This explains the special characteristic of GaN, whose GGA band gaps are even smaller than the LDA band gap, when calculated at the experimental lattice parameters. **Table I** lists the comprehensive calculation results for the seven semiconductors/insulators, where GaN is the only one that shows such abnormal behavior in general (for Si, only the PBEsol gap at experimental lattice is slightly lower than that of LDA).

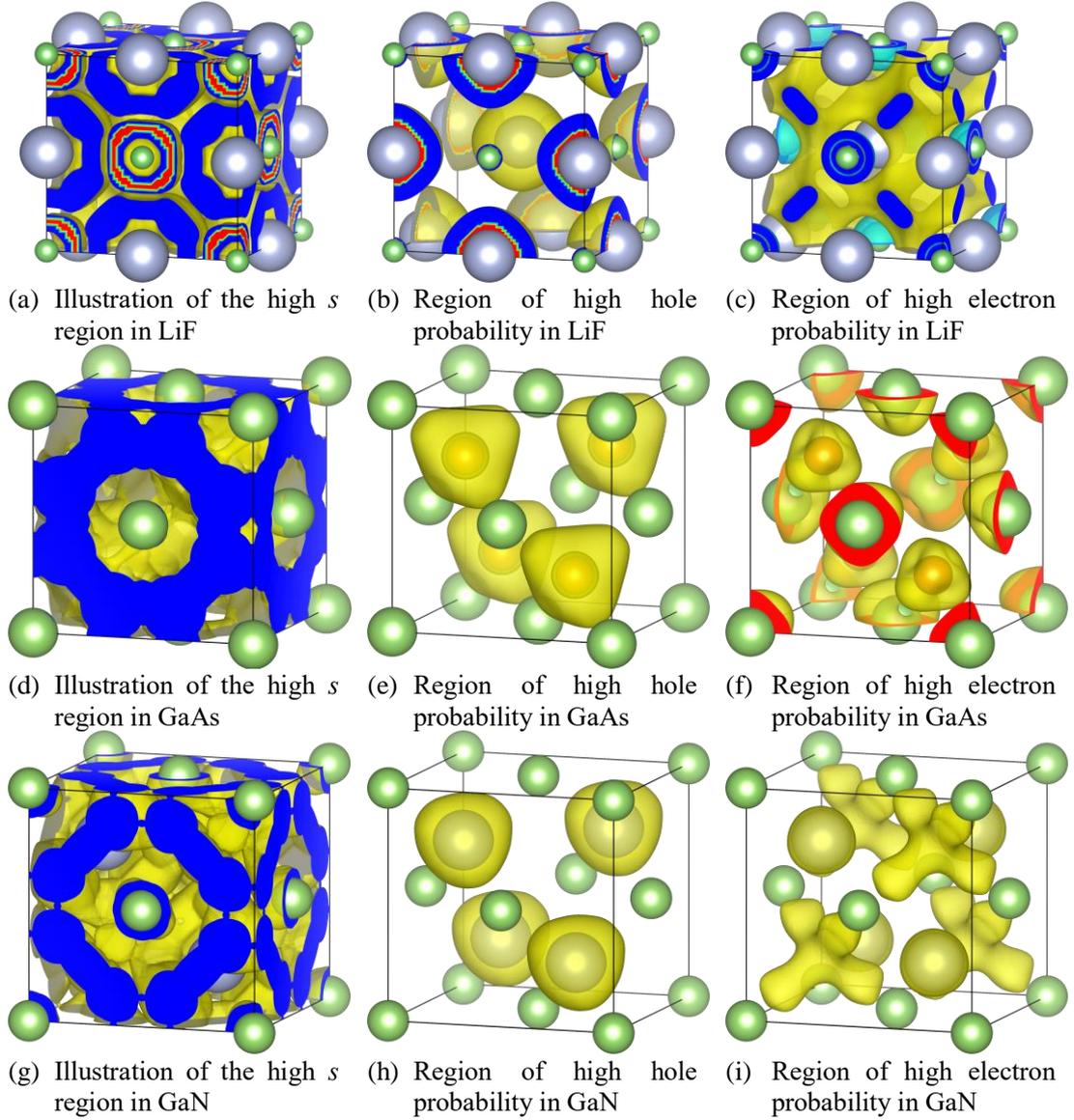

Figure 5. High s region as well as carrier distribution regions for several semiconductors: (a-c) LiF; (d-f) GaAs; (g-i) GaN.

## 3.2 ShGGA-1/2 versus shLDA-1/2

The primary concern of shDFT-1/2 is the amount of band gap aggrandizement ($\Delta E_g$) due to self-energy correction, compared with LDA/GGA

$$\Delta E_g = E_g(\text{shDFT-1/2}) - E_g(\text{LDA/GGA})$$

The final band gap predicted by shDFT-1/2 relies on both the fundamental LDA/GGA band structures, and also the degree of self-energy correction. The $E_g$ and $\Delta E_g$ values from self-consistent

shDFT-1/2 runs are also listed in **Table I**, with the optimized cutoff radii specified.

### 3.2.1 Relation of $\Delta E_g$ to lattice constant

The first observation is that compared with $E_g$, $\Delta E_g$ is much less dependent on the lattice constant, provided that the XC flavor and the charge stripping scheme in shDFT-1/2 are kept the same. This fact is observed by comparing the $\Delta E_g$ values from LDA/GGA-optimized lattice with $\Delta E_g$ values obtained under experimental lattice using the same XC. For instance, for Si one obtains dispersive $\Delta E_g$ values under optimized lattices among the four XCs, ranging from 0.735 eV to 0.811 eV. Nevertheless, for LDA only, $\Delta E_g$ is 0.735 eV under optimized lattice, but is kept almost the same (0.734 eV) when switched to experimental lattice. The same trend is observed for PBEsol, PBE and RPBE XCs as well.

For GGA XCs, $\Delta E_g$ is in general slightly higher when using experimental lattice constant, *i.e.*, the smaller lattice compared with the GGA-optimized lattice. Analysis on the self-energy potential cutoff radii reveals that the outer cutoff radius $r_{out}$ is typically diminished when switching to experimental lattice, which has strong correlation with the increase of $\Delta E_g$. At first glance, a shrink of $r_{out}$ should include less self-energy correction and yield lower band gaps, leading to contradiction. To understand this fact, we recall the variational process of optimizing $r_{out}$ and $r_{in}$ in shDFT-1/2. An optimal $r_{out}$ value should let the self-energy potential covering as much as the hole distribution region, but near $r_{out}$ there ought to be CB electron distribution region, otherwise $E_g$ will continue increasing rather than demonstrating a maximum. The constriction of lattice constant allows the hole to be localized in an even narrower region, and the reduced $r_{out}$ less disturbs the CB, explaining the enlarged $\Delta E_g$.

This phenomenon is consistent to the common rule that using a sharper cutoff, *i.e.*, higher power index $p$ in the self-energy potential cutoff function

$$\Theta(r) = \begin{cases} 0 & ,r < r_{in} \\ \left\{1 - \left[\frac{2(r - r_{in})}{r_{out} - r_{in}} - 1\right]^p\right\}^3 & ,r_{in} \leq r \leq r_{out} \\ 0 & ,r > r_{out} \end{cases}$$

renders a smaller $r_{out}$ ($r_{cut}$) and a higher $E_g$ in shDFT-1/2 (DFT-1/2). In **Figure 6**, the trend of $r_{in}$, $r_{out}$

and $E_g$ variations with respect to $p$ is illustrated for a typical ionic insulator LiF and a typical covalent semiconductor Ge.

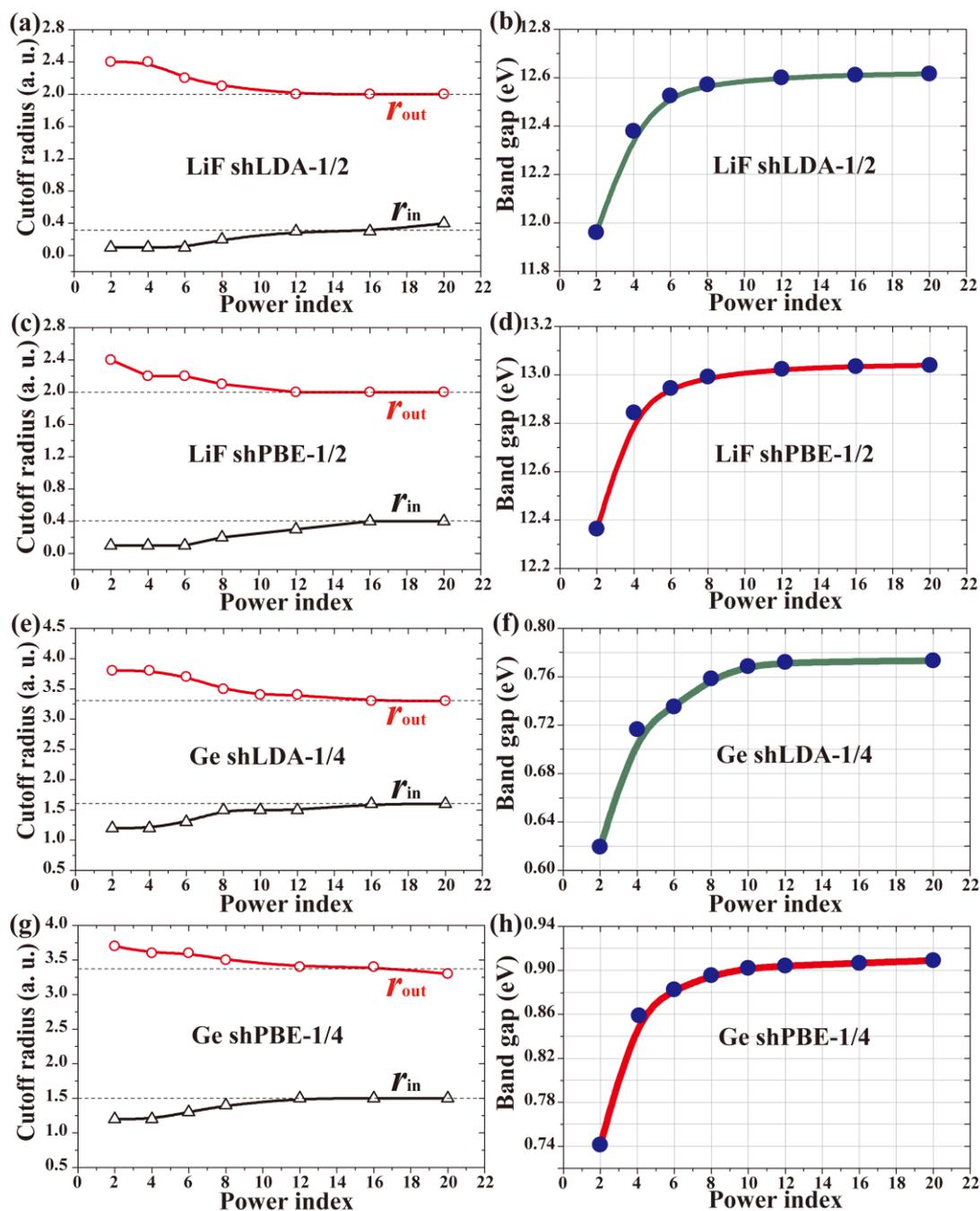

**Figure 6**. The power index ($p$)-dependence in typical shDFT-1/2 calculations. (a) Cutoff radii versus $p$ for LiF using LDA XC; (b) Band gap versus $p$ for LiF using LDA XC; (c) Cutoff radii versus $p$ for LiF using PBE XC; (d) Band gap versus $p$ for LiF using PBE XC; (e) Cutoff radii versus $p$ for Ge using LDA XC; (f) Band gap versus $p$ for Ge using LDA XC; (g) Cutoff radii versus $p$ for Ge using PBE XC; (h) Band gap versus $p$ for Ge using PBE XC.

### 3.2.2 Relation of $\Delta E_g$ to XC flavor

Another prominent rule is that, $\Delta E_g$ for GGA is universally larger than that of LDA, both at equilibrium lattice constant and at experimental lattice constant. One may suspect that the GGA self-energy potentials are stronger than that of LDA. To clarify this point, we plot several typical self-energy potentials in **Figure 7**. The horizontal axis is the radius $r$ measured from the center of the atom that is subject to self-energy correction, while the vertical axis is the product of $r$ and the corresponding self-energy potential $V_S(r)$. For convenience and consistency, even for Ge and As we show the normal -1/2 self-energy potentials, despite the fact that the actually used ones are -1/4 for Ge and GaAs. According to Gauss' law, $rV_S$ has to converge to -0.5 for infinitely large $r$. For small radius $r$ (within 1 bohr or 2 bohr), the discrepancy of self-energy potentials among different XCs is very minor. On the other hand, optimized $r_{out}$ are usually smaller than ~4 bohr. Hence, only the critical $r$ regions are demonstrated for clearer comparison.

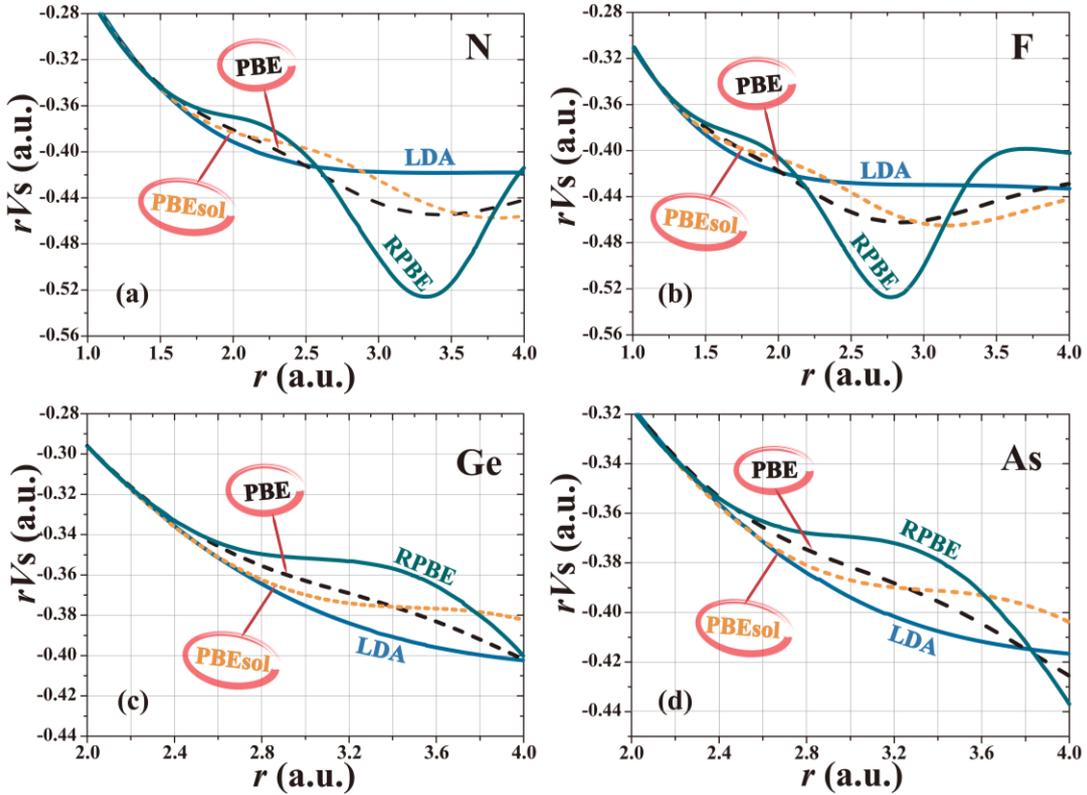

**Figure 7**. Self-energy potentials of various XC flavors for (a) N; (b) F; (c) Ge; (d) As.

For N, the self-energy potential of LDA is even stronger than the three selected GGA ones, for the radius $r$ less than ~2.5 bohr, which is the optimal $r_{out}$ of GaN under all XCs. The optimized $r_{out}$ for F is merely 2.0 bohr in LiF, but for the region of 0–2 bohr, the self energy potential of LDA flavor is even stronger than that of PBEsol, PBE and revPBE. The self-energy potentials of Ge and As demonstrate the similar trend, with $V_S$ under LDA being the strongest for the principal range of $r$. These facts reveal that the enhanced $\Delta E_g$ values in GGA are not simply due to stronger self-energy potentials.

Therefore, we examined another possible factor that influences $\Delta E_g$, the self-consistent shDFT-1/2 process-induced charge density variation. As is well-known, the original and standard DFT-1/2 method requires self-consistency between the wavefunction and charge density under the influence of foreign -1/2 potential. The spatial region of hole localization will become lower-energy regions, thus more valence electrons transfer to this region, yielding an additional backward electric field that can pull up the VB a little bit. For this reason, the non-self-consistent DFT-1/2 band gaps are consistently larger than self-consistent DFT-1/2 band gaps [20]. For convenience, we define

$$E_{\text{shift}} = E_g(\text{non-self-consistent}) - E_g(\text{self-consistent})$$

which is normally positive. For all materials under investigation, their $E_{\text{shift}}$ values are illustrated in **Figure 8**. A clear trend is discovered that for high-$F_X$ XCs such as RPBE and PBE, $E_{\text{shift}}$ becomes minor. LDA, with no gradient-induced exchange enhancement on the other hand, presents the largest difference between non-self-consistent and self-consistent shDFT-1/2 band gap values. A rather surprising discovery is that, $E_{\text{shift}}$ even becomes negative under the RPBE functional for Ge. However, the trend of charge variation when switching on self-consistent shDFT-1/2 runs, as illustrated in **Figure 9**, confirms that more electrons tends to enter the self-energy potential region for *all* compounds, including Ge.

We give a rough explanation to such abnormal phenomenon in Ge. Note that its VB electrons are localized around the Ge-Ge bond centers (referred to as the VB region) [11], and a density gradient exists from the outer part pointing to the VB region. The trimmed self-energy potential, as it lowers the electronic energy levels, attracts more electrons to enter the VB region, which enlarges $|\vec{\nabla} n|$ around the VB region. As $n$ is already large there, the charge transfer process due to the self-consistency in shDFT-1/2 does not alter $n^{4/3}$ by a similar large extent like $|\vec{\nabla} n|$. Consequently, such charge transfer should enhance the *s* values around the VB region, leading to a downshift of VB that is specific to GGA. As $F_X$ increases sharply with respect to $s$ for RPBE (**Figure 1**), such effect may even overwhelm the backward electric field, predicting a higher $E_g$ value under the RPBE XC.

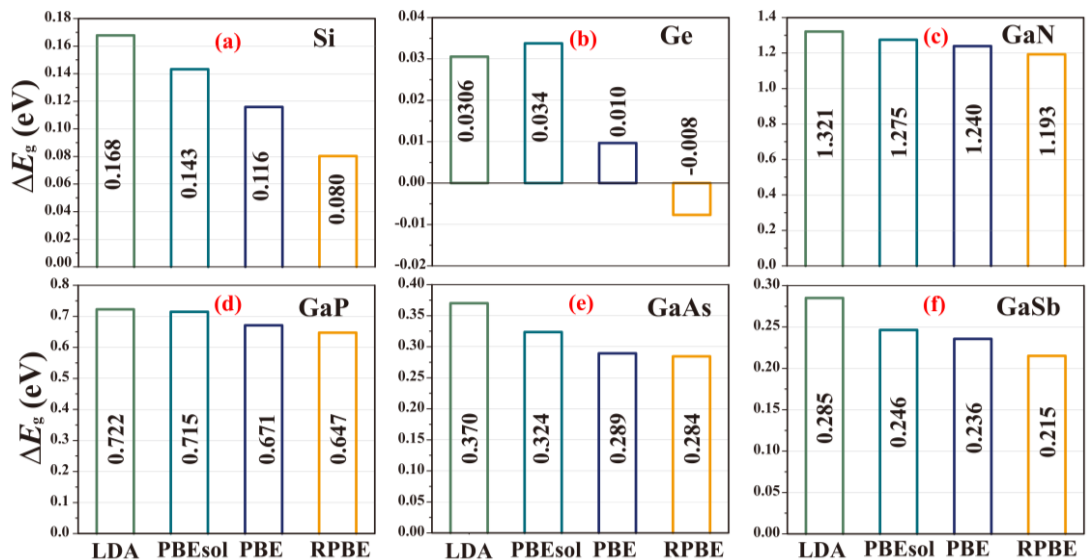

**Figure 8**. Band gap difference between non-self-consistent and self-consistent shDFT-1/2 results for selected semiconductors.

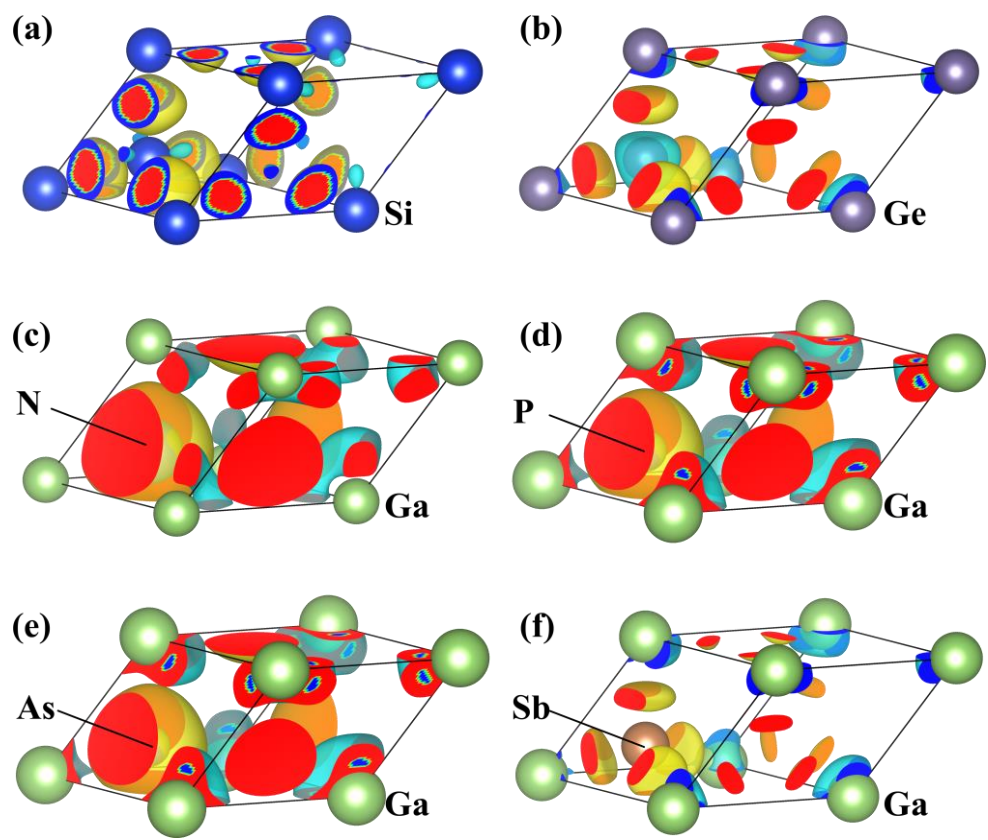

**Figure 9**. Charge density difference between non-self-consistent and self-consistent shLDA-1/2 results for selected semiconductors: (a) Si; (b) Ge; (c) GaN; (d) GaP; (e) GaAs; (f) GaSb. Yellow

and cyan regions enclose positive (electron net input region in self-consistent run) and negative regions. Contour densities: (a) ±0.01 e Å$^{-3}$; (b-f) ±0.02 e Å$^{-3}$.

**3.2.3 Additional discussion**

It has been established that for wide gap ionic insulators ($E_g$ > 6 eV), both self-consistent LDA-1/2 and self-consistent GGA-1/2 predict lower band gaps than experimental [20]. For medium gap ionic insulators (3 eV < $E_g$ < 6 eV), self-consistent GGA-1/2, especially using the PBEsol XC, is quite satisfactory in terms of band gap prediction [15]. When it comes to typical covalent semiconductors, where non-self-consistent runs are improper even from theoretical [20], shGGA-1/2 will be evaluated against shLDA-1/2 as follows.

Among the six covalent semiconductors under investigation (LiF not included as it is a highly ionic insulator), shLDA-1/2 in general shows better performance in terms of the band gap, for calculations under both optimized (@opt) and experimental (@exp) lattice constants. Compared with experimental gap, the mean errors are: 0.130 eV for LDA@opt, 0.045 eV for PBEsol@opt, 0.061 eV for PBE@opt, 0.105 eV for RPBE@opt; 0.037 eV for LDA@exp, 0.044 eV for PBEsol@exp, 0.236 eV for PBE@exp, 0.298 eV for RPBE@exp. The constantly positive values reflect the general trend of shDFT-1/2 to predict slightly larger band gaps than experimental. On the other hand, the mean absolute errors are: 0.132 eV for LDA@opt, 0.150 eV for PBEsol@opt, 0.254 eV for PBE@opt, 0.418 eV for RPBE@opt; 0.203 eV for LDA@exp, 0.229 eV for PBEsol@exp, 0.236 eV for PBE@exp, 0.298 eV for RPBE@exp. When fixed to experimental lattice, GGA already yields larger band gaps than LDA. The shGGA-1/2 gaps are even larger, probably unsuitable for engineering applications. Theoretically, the quasi-particle band gaps predicted by self-consistent *GW* are also typically over-large [35]. However, in shGGA-1/2 calculations, fixing to experimental lattice constants is not suitable to reach engineering accuracy. It is better to use equilibrium lattice constants for shGGA-1/2, while in shLDA-1/2 calculations it is reasonable to use experimental lattice constant to overcome the shortcoming of over-binding in LDA, though the overall band gap quality of shLDA-1/2 at optimized lattice constants is also supreme.

As a final remark, shDFT-1/2 enables the investigation of calculated band gap versus lattice constants for those narrow gap semiconductors that are predicted to be metallic by LDA/GGA. Typical examples include Ge and GaSb. For Ge, a maximum gap is also discovered. Yet, the experimental lattice constant of Ge falls into the downhill side, which is opposite to the case of Si. The same phenomenon is discovered for GaSb. **Figure 10** illustrates the shDFT-1/2 results for Ge, GaSb and Si, which again confirm the special characteristic of Si.

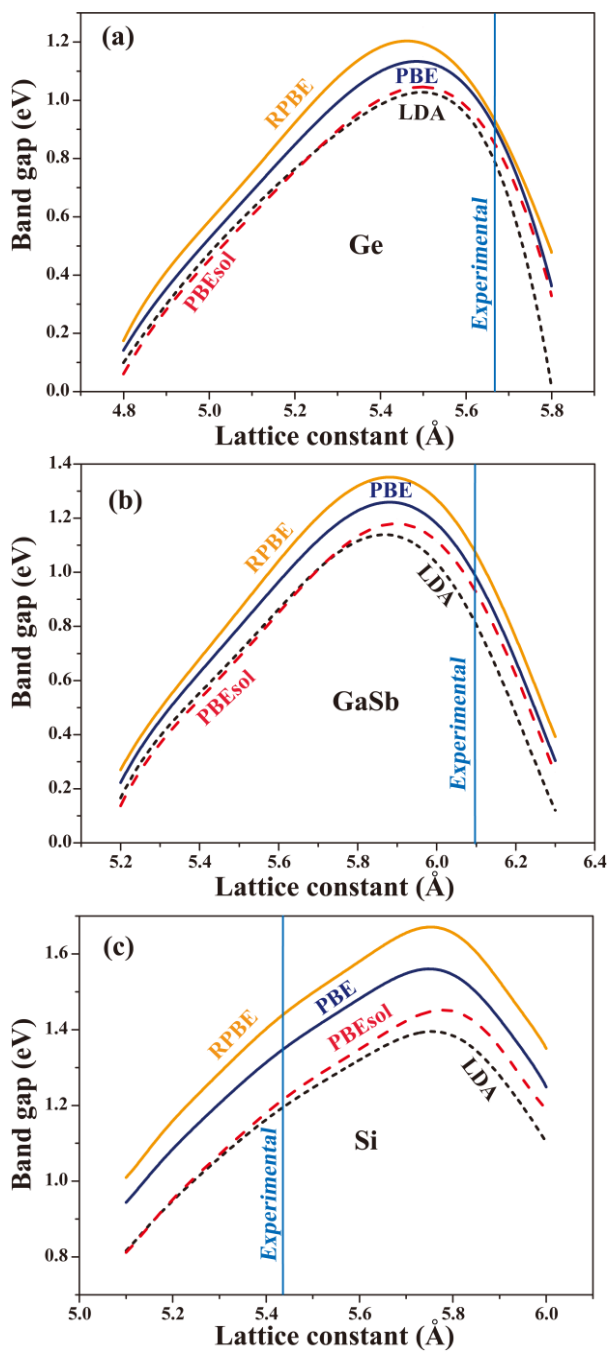

**Figure 10**. Impact of lattice constant on the shDFT-1/2 band gaps for (a) Ge and (b) GaSb, with comparison to (c) Si.

# 4. Conclusion

The shDFT-1/2 band gaps have been extensively tested using LDA and GGA (PBEsol, PBE, RPBE) XCs for typical semiconductors Si, Ge, GaN, GaAs, GaP as well as GaSb. The following conclusions are drawn towards the impact of GGA and the accuracy of shGGA-1/2.

(1) When calculated at optimized lattice parameters, the most significant impact of XC flavor lies in the magnitude of lattice constant. Most semiconductors show a reduced band gap upon enlarging the cell volume, in a considerable region around the experimental volume, but some indirect gap semiconductors like Si and GaP may demonstrate an opposite trend, which comes from the type change between indirect and direct gaps. The basic LDA/GGA gap depends greatly on the lattice constant, and the self-energy correction-induced gap enhancement ($\Delta E_g$) is relatively insensitive to lattice constant. Hence, GGA/shGGA-1/2 band gaps are usually smaller than that of LDA/shLDA-1/2 (with the exception of Si), due to the dominating effect of lattice constant-dependency.

(2) When evaluated at experimental lattice parameters, shGGA-1/2 band gaps are larger than shLDA-1/2, with the only exception of GaN. The reason lies in that the valence band states are usually more localized than the conduction band states, thus the same exchange enhancement factor from GGA tends to yield more absolute downshift of the valence band than the conduction band. GaN is special in that its conduction band states extend to the strong gradient-correction region, leading to a severe downshift of the conduction band.

(3) For typical covalent semiconductors, in the average sense shGGA-1/2 predicts a bit low band gaps at the optimized lattice parameters, but yields too large band gaps if fixed to experimental lattice. And under GGA it is recommended to use optimized lattice parameters because the over-large cell volume-induced gap reduction can be compensated to some extent by the exchange enhancement-induced gap widening. For shLDA-1/2, on the other hand, calculations at both optimized and experimental lattice parameters consistently yield satisfactory band gap values.

# Acknowledgement

This work was supported by the National Natural Science Foundation of China (61974049).